# Effects of the Number of Active Receiver Channels on the Sensitivity of a Reflector Antenna System with a Multi-Beam Wideband Phased Array Feed

Oleg Iupikov

Sevastopol National Technical University, Ukraine

*Abstract* — A method for modeling a reflector antenna system with a wideband phased array feed is presented and used to study the effects of the number of active antenna elements and associated receiving channels on the sensitivity of the system. Numerical results are shown for a practical system named APERTIF that is currently under developed at The Netherlands Institute for Radio Astronomy (ASTRON).

*Index Terms* — phased array feeds, antenna system sensitivity, reflector antennas.

## Introduction

At present, several types of so called 'dense' Phased Array Feed (PAF) systems for reflector antennas have been considered for application in future radio telescopes [1—6]. The main advantage of these PAFs over conventional horn feeds is that PAFs can be equipped with digital beamformers (providing individual complex control per element) and realize an optimal illumination of the reflector aperture with almost complete cancellation of the spillover losses. Another advantage is that the inter-element separation distance is much smaller than a wavelength that allows forming multiple closely overlapping beams and hence continuous Field Of View of deep reflector antennas ( with the focal ration $F/D < 1$). Such a wide uniform FOV cannot be realized with conventional clusters of horns [7]. On the other hand, the elements of PAFs are strongly coupled with a consequence that the low noise amplifiers connected to these elements will be 'coupled' as well. If the PAF elements are not optimally noise matched to the LNAs, this can lead to significant increase of the total system noise temperature ($T_{sys}$) [8]. The optimal impedance noise match condition requires the active reflection coefficients of the PAF elements to be equal to the optimal reflection coefficient of the LNAs that, in practice, is difficult to realized since the active reflection coefficient is dependent on the beam former weights and can be different for different elements. In the articles [8, 9] the theory of optimal impedance and noise matching is described. It was used in the current work to compute noise parameters of the antenna system. Obviously, the larger the size of the arra , the more multiple beams (wider FOV) can be formed. However, since only a relatively small fraction of the array is 'actively' used in beamforming of each beam (meaning that only a relatively small number of elements have high weights), the relative system cost (dominated by the cost of the beamformer and required processing) is increasing in the same time.

The goal of the current work is to analyze the effects of reducing the number of actively used receiver channels on the key performance parameters of the system. The "reciever channel" means here the chain "antenna element of the array — low-noise amplifier (LNA) — cable — reciever".

One of the most important parameters of radio telescopes is sensitivity since the observation speed is proportional to the sqared sensitivity [10]. That's why the sensitivity was choosen as the system quality factor. It is assumed that maximum allowed sensitivity reduction can be 5 % with respect to the sensitivity which can be realized with the array in which all elements are used in beamforming.

Using the showed below analysis one can find a trade-off solution between the maximum system sensitivity and the minimum number of active reciever channels (system cost).

The analysis has been carried out for an example of the APERTIF system [11] wich is currently under developing in the Netherlands Institute for Radio Astronomy (ASTRON).

**1. Problem formulation**

Each reflector antenna of the interferometer consists of the 25-m parabolic reflector with the antenna array of 100+ antenna elements that is located in the reflector focal plane. The signals received by each element of the array is amplified by low noise amplifiers (LNAs). The beamforming is realized in the digital domain after frequency channelization and filtering. For detailed description of the APERTIF system see e.g. [11]. The key question of this study is how many receiver channels can be excluded *from beam forming process* with the maximum allowed system sensitivity reduction of 5%. Note that we are not excluding array elements physically. We can't do it since i) active part of the array should remain dense (distance between array elements is less then λ/2) to form required beam correctly and ii) maximum scan angle depends on physical size of the array.

**2. Algorithm of the analysis**

To solve this task the following algorithm of analysis can be used:

Step 1. Electromagnetic (EM) simulation of the PAF. At this step the current distribution in metal structure of the array as well as impedance matrix and radiation embedded element patterns (EEP) are computed.

Step 2. Computation of the secondary EEP (patterns after reflection from the dish)

Step 3. Microwave (MW) simulation of the whole system (see paragraph 2.3 below) and computation of the optimal weight coefficients. The maximum SNR beamforming was used in this study.

Step 4. Make zero those weight coefficients which are below threshold $\Delta W_{trshld}$. Threshold $\Delta W_{trshld}$ is varying in range -30…0 dB (see step 6).

Step 5. Compute combined pattern of the array, combined pattern, noise parameters and sensitivity of the whole system for computed in steps 3–4 weights.

Step 6. Repeat steps 4–5 for the range of $\Delta W_{trshld}$ from -30 to 0 dB.

Let us describe each step in details.

**2.1. Electromagnetic simulation**

The current distribution in the metal structure of the array was computed by using Characteristic Basis Function Method (CBFM) [12—14]. This method is extension of the Method of Moments (MoM) and it is developed for modeling finite pseudo periodic structures. It allows computation the current distribution in the array but has much lower requirements for the computer powers then MoM.

CBFM allows analyzing big metal structures that impossible (at present days) to analyze by MoM, but that are still too small to use analysis of infinite arrays [15, 16].

As result of EM simulation we have impedance matrix of the array **Z** and complex EEPs.

### 2.2. Simulations of the secondary field patterns

The secondary patterns were computed in the program GRASP9, in which method of Physical Optics (PO) and Geometrical Theory of Diffraction (GTD) are implemented [17].

### 2.3. Modeling of the antenna-receiver system and determination of the optimal weights computation

The metal structure of the array can be thought of as *N*-port microwave device that is generating signal **e** and noise **c** waves when receiving the signal. The microsrtip feed [18] and LNA were connected to this device and MW simulation was performed [19].

To outline the optimization procedure with PAF beamformers as considered in this paper, we utilize the generalized system representation as shown in Fig. 1 for the *N* actively beamformed antennas. The system is subdivided into two blocks: (i) the frontend including the reflector, array feed, LNAs; and (ii) the beamformer with complex conjugated weights $w_n$ and an ideal (noiseless/reflectionless) power combiner realized in software. Here, $\mathbf{w}^H = \begin{pmatrix} w_1^* & w_2^* & \ldots & w_N^* \end{pmatrix}$ is the beamformer weight vector, *H* is the Hermitian transpose, and the asterisk denotes the complex conjugate. Furthermore, $a = \begin{pmatrix} a_1 & a_2 & \ldots & a_N \end{pmatrix}^T$ is the vector holding the transmission-line voltage-wave amplitudes at the beamformer input (the *N* LNAs outputs).

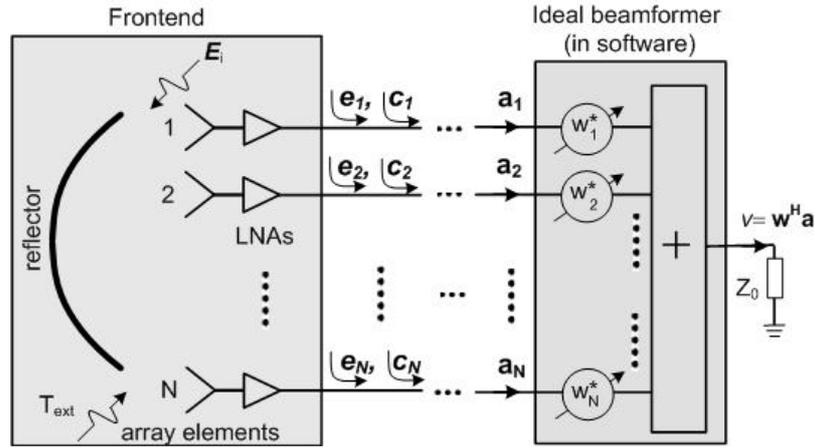

Fig. 1. Structure of the antenna system

Frontend is characterizing by noise-wave correlation matrix

$$\mathbf{C} = \begin{pmatrix} c_{11} & c_{12} & \cdots & c_{1N} \\ c_{21} & c_{22} & \cdots & \vdots \\ \vdots & \vdots & \ddots & \vdots \\ c_{N1} & \cdots & \cdots & c_{NN} \end{pmatrix}$$

and by signal vector **e**, elements of which are signal waves [15, p. 163-166] on the outputs of each array antenna element (noiseless) when receiving the signal:

$$\mathbf{e} = \begin{pmatrix} e_1 & e_2 & \cdots & e_N \end{pmatrix}^T.$$

There are many beamforming schemes exist to compute weight coefficients. In presented study the Maximum SNR beamforming [19] was used:

$$\mathbf{w} = \mathbf{C}^{-1}\mathbf{e},$$

in which the vector **e** describes signals when incident plane wave is coming from direction of interest.

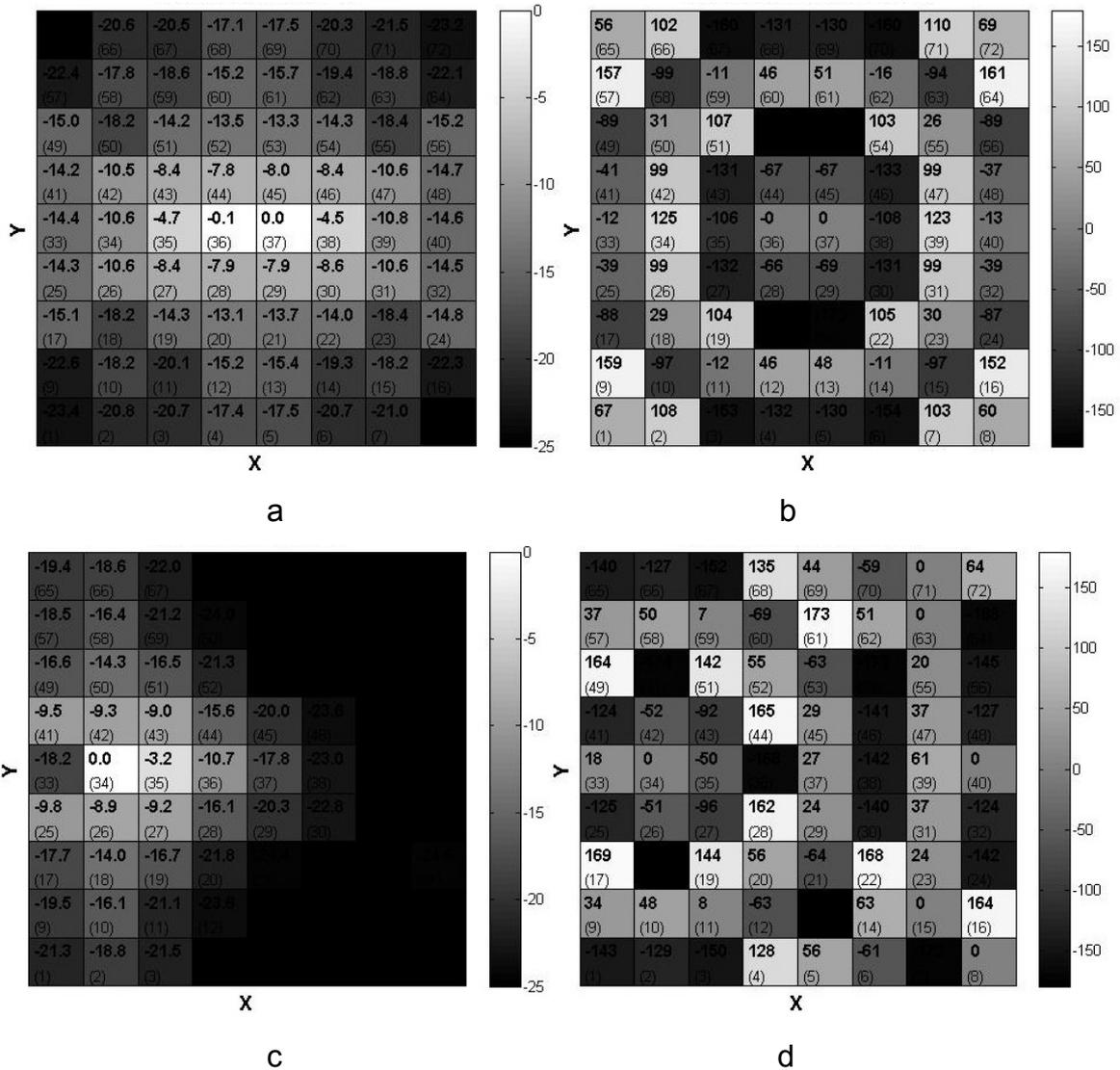

Fig. 2. The array weight coefficients for the on-axis (a, b) and off-axis (c, d) beams: absolute values in dB (a, c) and phase in degrees (b, d)

Since the antenna array is focal plane array the main power of the incident plane wave is focused on several array elements, not on the all of them. Fig. 2 shows normalized absolute value of the weight coefficients (MaxSNR beamforming) for the array of Vivaldi antenna elements 8x9 at 1420 MHz while scanning angles are $\theta=0°$ (fig. 2a,b) and $\theta=1,5°$ (fig. 2c,d). At the figure the rectangles denote array antenna elements of the same polarization at the position in the array.

As one can see at the figure, not more than half of the array elements are strong excited ($20\lg(|w|) > -10$ dB). Therefore, not all elements can be used to form the beam. It will reduce the number of receiver channels and receiver and cables cost. However, the high frequency low-loss switchers should be used after LNA to switch subarrays.

**2.4. System sensitivity for different values of the threshold $\Delta W_{trshld}$**

After EEPs and weight coefficients had been computed the weight coefficients below $\Delta W_{trshld}$ dB were set to zero (step 4 of the algorithm). For the resulting set of weights the combining radiating pattern, system noise temperature and sensitivity were computed (step 5 of the algorithm). All antenna elements (both active and inactive) were loaded by LNA.

The beam sensitivity, aperture efficiency, system noise temperature and spillover efficiency as functions of threshold $\Delta W_{trshld}$ are shown at Fig. 3. Computations were done for on-axis beam and off-axis beam scanning at 1.5 deg.

As one can see, for the considered system to form on-axis beam it is enough to use only 45 receiving channels instead of 72 (number of antenna elements of the same polarization) with the sensitivity reduction less than 5 %. For the off-axis beam the number of active receiving channels can be even less since large part of receiving energy is not caught by array (this fact one can see at the Fig. 2c). This means that the number of the active receiving channels can be reduced up to 40 % with the sensitivity reduction less than 5 %. However, the subarrays must be switched to form off-axis beams.

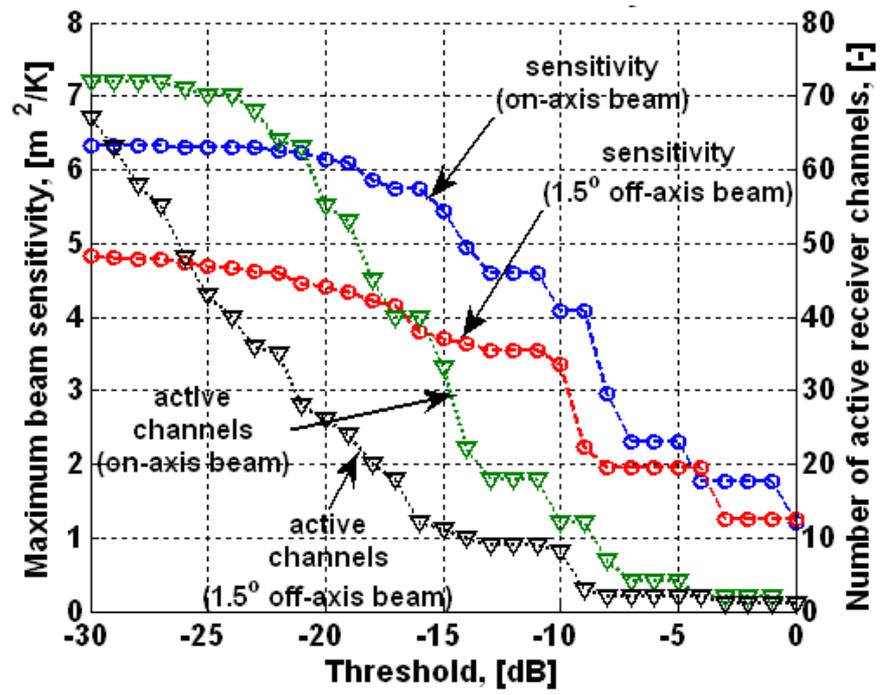

a

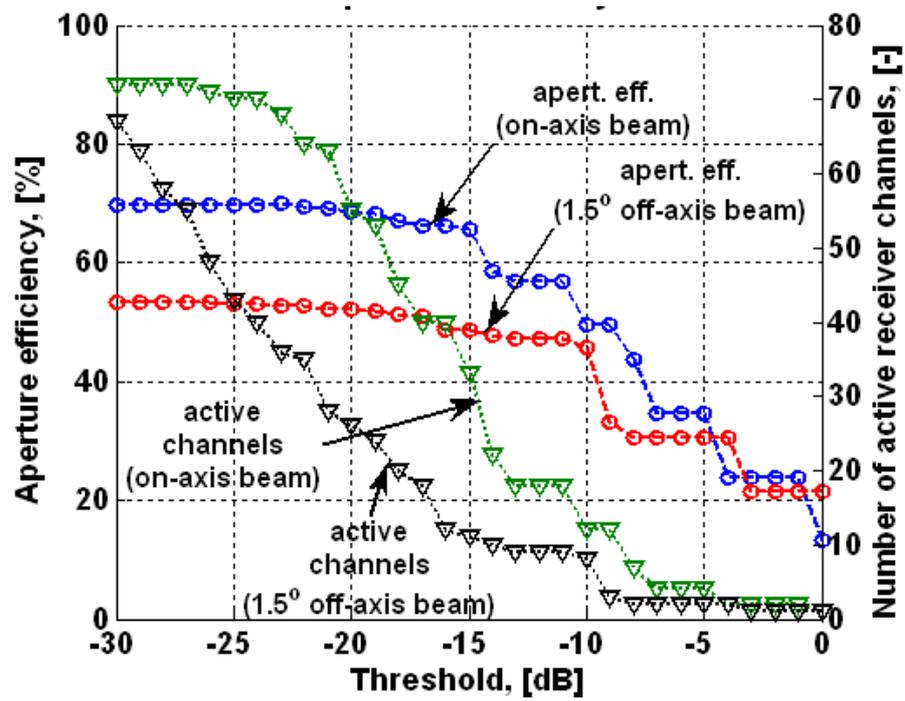

b

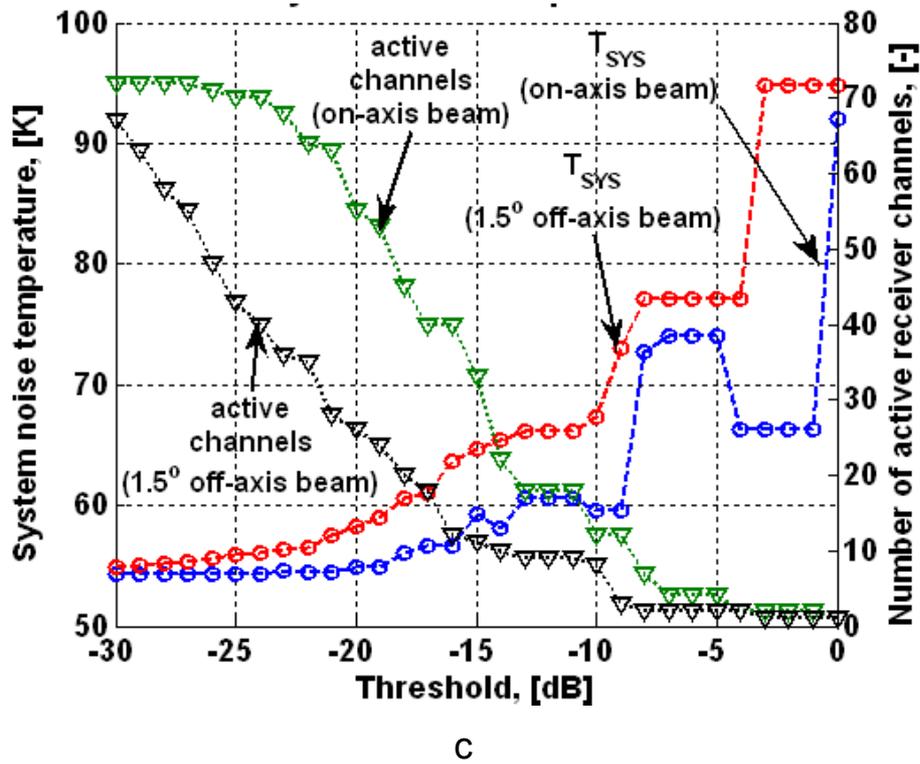

c

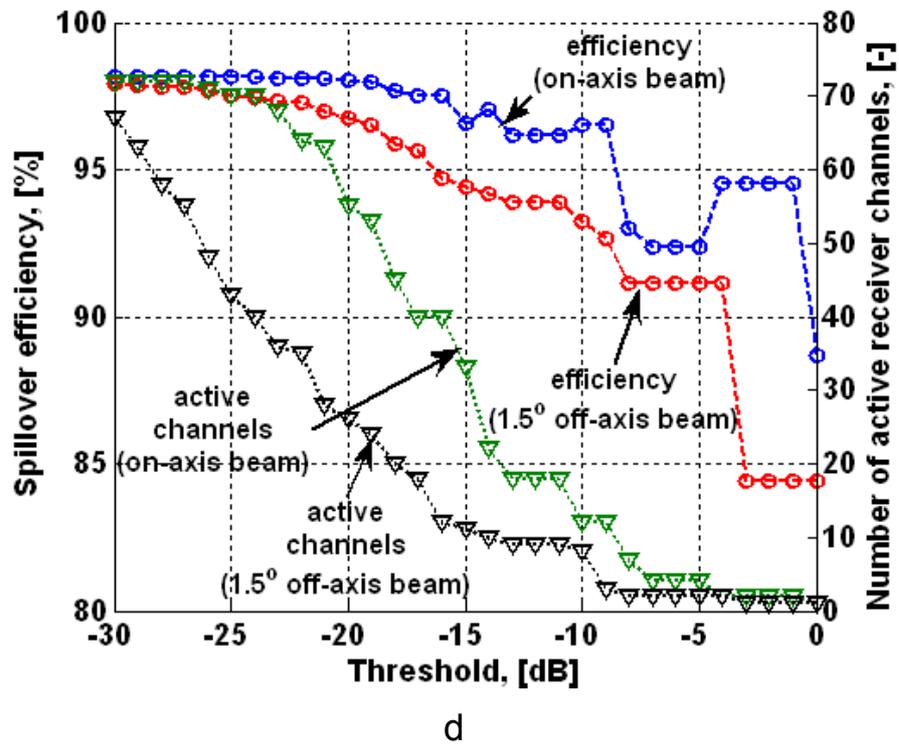

d

Fig.3. System parameters and number of active receiving channels as functions of threshold $\Delta W_{trshld}$ for the on-axis and off-axis beams: (a) beam sensitivity, (b) aperture efficiency, (c) system noise temperature, (d) spillover efficiency.